\documentclass[sigconf]{acmart}
\renewcommand\footnotetextcopyrightpermission[1]{}

\usepackage{booktabs} 
\usepackage{subfigure}
\usepackage{multirow}
\usepackage{url}
\usepackage{enumerate}
\usepackage{colortbl}
\usepackage{float}
\setcopyright{none}

\begin{document}
\title{Who Will You Share a Ride With: Factors that Influence Trust of Potential Rideshare Partners}

\author{Yang Zhou}

\affiliation{%
  \institution{Computer Sicience and Engineering Department}
  \streetaddress{University of North Texas}
  \city{Denton} 
  \state{Texas} 
  \postcode{76207}
  \country{USA}
}
\email{yangzhou2@my.unt.edu}

\author{Yan Huang}
\affiliation{%
  \institution{Computer Sicience Engineering Department}
  \streetaddress{University of North Texas}
  \city{Denton} 
  \state{Texas} 
  \postcode{76207}
  \country{USA}
}
\email{Yan.Huang@unt.edu}

\author{Joseph McGlynn}

\affiliation{%
  \institution{Communication Studies Department}
  \streetaddress{University of North Texas}
  \city{Denton} 
  \state{Texas} 
  \postcode{76203}
  \country{USA}}
\email{Joseph.Mcglynn@unt.edu}

\author{Alexander Han}
\affiliation{%
  \institution{Flower Mound High School}
  \city{Flower Mound} 
  \state{Texas} 
  \postcode{75022}
  \country{USA}
}
\email{alex.han@gmail.com}


\begin{abstract}
Ridesharing can save energy, reduce pollution, and alleviate congestion. Ridesharing systems allow peer-to-peer ride matching through mobile-phone applications. In many systems people have little information about the driver and co-riders. Trust heavily influences the usage of a ridesharing system, however, work in this area is limited. In this paper, we investigate the role of social media in shaping trust and delineate factors that influence the decisions of trust placed on a driver or co-rider. We design a ridesharing system where a rider can see the Instagram profile and 10 recent pictures of a co-rider and ask them to make judgments on six trust related questions. We discovered three important factors that shape trust forming: \textit{social proof}, \textit{social approval}, and \textit{self-disclosure}. These three factors account for 40\%, 33\%, and 13\% of the variance, and subsequent factors extracted accounted for substantially less variance. Using these three factors, we successfully build a trust prediction model with F-measure of 87.4\%, an improvement from 78.6\% on using all factors collected. 
\end{abstract}

%
%

\keywords{ Ridesharing, social network, Instagram, Exploratory Factor Analysis, logistic regression }

\maketitle

\let\thefootnote\relax\footnote{Copyright is held by the author/owner(s).}
\let\thefootnote\relax\footnote{UrbComp'17, August 14, 2017, Halifax, Nova Scotia, Canada.}

\section{Introduction}
Urban traffic gridlock is a familiar scene. At the same time, the mean occupancy rate of personal vehicle trips in the United States is only 1.6 persons per vehicle mile. Ridesharing can solve many environmental, congestion, and energy problems. With these advantages, ridesharing is becoming increasingly popular. Many ridesharing services such as Uber and Lyft allow for peer-to-peer ride matching through mobile-phone application \cite{ huang2014large}. However, with current ridesharing systems, a user has little information about their rideshare partner. A user cannot choose or approve a partner before the ride. There are risks and social discomfort that arise due to lack of trust amongst co-passengers. Trust in co-passengers is the most important factor influencing willingness to give or take rides from others \cite{chaube2010leveraging}. Understanding the factors that influence perceptions of trust among strangers has important implications 
for research seeking to reduce urban gridlock. Ridesharing has great potential to increase carpooling and reduce urban congestion, but only if the users trust the potential rideshare partner and do not feel they are entering a risky situation.

With the widespread use of social media, it is possible to obtain basic information of users who want to use a ride-sharing service. Social media allows people to create, access, and interact with a wide range of information \cite{jang2015generation}. In this work, we use Instagram to extract trust factors. Instagram allows users to share photos and comments. It has quickly become one of the most popular social media platforms in the United States since its launch in 2010. By the end of January 2017, Instagram had 600 million active users per month with 400 million of these users accessing the platform every single day \cite{Instagram}. Instagram has two types of users, private-profile users and public-profile users. With public-profile users, the profiles are available for viewing even if one does not follow a user. With private-profile users, one must get the permission of the user to follow in order to see their posts. Unlike other social media, each post by an Instagram user is a picture or a video \cite{ hu2014we}. 

By combining ridesharing systems with social media, we attempt to provide more information for assigning trust on a rider or driver in a ridesharing system. We provide pictures and useful social information from a user$'$s Instagram.  Our analysis identifies three key factors that influence trust formation. By identifying key factors that people use when making assessments of trustworthiness for social media users, understanding is increased regarding what aspects of a profile compel users to increase perceptions of trust and reduce perceptions of risk. This paper makes the following contributions:

\begin{itemize}
\item We propose a system that allows a user to gain more information of a potential co-rider through linking social media and ridesharing system.

\item We identify three crucial factors that influence trust building:\textit{ social proof}, \textit{social approval}, and \textit{self-disclosure} from social media posts. These three factors account for 40\%, 33\%, and 13\% of the variance, respectively. Subsequent factors extracted accounted for substantially less variance. This result expands our understanding of which factors persuade people to trust potential rideshare partners as co-riders. In addition, we analyze the manipulability of the factors and shed light on the robustness of the information that could be used to build trust. 

\item As researchers and practitioners gain understanding of factors that influence perceptions of social media trust, increased collaboration can occur among users and rideshare companies to provide the information that users need to view for adequate levels of trust to be reached.

\item Using three key factors, we successfully build a trust prediction model with F measure of 87.4\%, an improvement from 78.6\% on using all factors collected. 
\end{itemize}

By examining the factors of user profiles that lead to perceptions of increased trust and decreased risk, these results provide insights for both consumers and organizations on the information needed to increase participation in ridesharing applications through increased trust and reduced perceptions of risk.

\section{Related Work}
Vineeta et al. \cite{chaube2010leveraging} summarized the results of a rideshare need assessment survey conducted within the Virginia Tech University. The purpose of this work is to understand commuter travel patterns, their needs, and to identify their preferences for private vehicles and public transit across a variety of travel contexts. The results show that although people seem to be interested in ridesharing, it remains an unpopular mode of transportation. The most important reasons for its unpopularity are the inconvenience, lack of trust, and incentives. Among these three factors, trust is a major factor that people consider while offering or accepting a ride. 

Other systems also find trust as one of the major concerns of rideshare users. Joshua et al. \cite{morse2007carloop} found that people are often hesitant to carpool with unknown drivers. They developed a carpooling system which connects commuters in the same organization with each other, leveraging the common background of their shared employer. Rick et al. \cite{wash2005design} designed the RideNow project to help individuals within a group or organization to coordinate ad hoc shared rides. RideNow service is intended to coordinate ride sharing within groups or small organizations of a few hundred to a few thousand people. RideNow project can reduce the uncertainty of dealing with strangers and has been found to increase conversations and dialogue among riders. These works benefit users within the same community and are not designed to address trust issues in general ridesharing.

Zimride \cite{zimride} is one of the applications that combines ridesharing with social network sites such as Facebook and MySpace. Zimride can deal with trust problem in rideshare systems. For instance, a Myspace user can request a ride or offer a ride to other members of their corporation or university. In this approach, users who will share a ride together have a common ground.

Both Uber and Lyft use a rating system. At the end of each ride, a rider can rate a driver and also give feedback based on the driver$'$s driving skill, behavior, and condition of the vehicle. If the cumulative rating is below the minimum for the driver's area, the driver is suspended \cite{Uber}. On the other hand, a driver also has the option to rate a rider based on cancellations, no-shows, and late arrivals. A user$'$s rating will affect a driver$'$s willingness to pick up the user. A rating-based system suffers from a cold start problem when insufficient ratings are available. Ratings can also be biased, as not all users give ratings, especially when the number of ratings of a driver are small. Often, people who are extremely unhappy tend to give ratings with greater frequency.

\section{Data Collection}
We collected the data on January 20, 2017 from Instagram. We first chose one seed public user and then crawled all the followers of the seed user. We chose the public user UNT as our seed. There were a total 27k followers of UNT by the end of the day. Among those users, we filtered out private users, resulting in 8,436 public users. We then chose 100 users randomly from these public users. 

\subsection{Attributes of Dataset}
The dataset of Instagram includes various pieces of a user$'$s information. We summarized the information into three basic types: user information, user account information, and post information. There are several attributes under these three types. The detailed information is listed in Table 1 to Table 3.

\begin{table}[!ht]
\centering
\caption{User information}
\begin{tabular}{|c|c|} \hline
\textbf{\textbf{User information}}&\textbf{}\\ \hline \hline
\texttt{\textbf{User name}} &Instagram account name, user can change it\\ 
\texttt{}&  at anytime.\\ 
\hline
\texttt{}& A user can type their true name or other\\ 
\texttt{\textbf{Full name}} & preferred name. If a user did not enter the \\ 
\texttt{} & full name, it will not display the full name \\ 
\texttt{} & part of this user \\ 
 \hline
\texttt{\textbf{Biography}}& A user can type a few sentences to describe\\ 
\texttt{} & himself. \\ 
\hline
\end{tabular}
\end{table}

\begin{table}[!ht]
\centering
\caption{User account information}
\begin{tabular}{|c|c|} \hline
\textbf{\textbf{User account information}}&\textbf{}\\ \hline \hline
\texttt{\textbf{Followers}} & Total number of users who are\\ 
\texttt{}& following this user.\\ 
\hline
\texttt{\textbf{Following}}& Total number of users that this\\ 
\texttt{} & user is following. \\ 
 \hline
\texttt{\textbf{Posts}}& Total number of posts that the\\ 
\texttt{} & user has posted through Instagram.\\ 
\hline
\end{tabular}
\end{table}

\begin{table}
\centering
\caption{Post information}
\begin{tabular}{|c|c|} \hline
\textbf{\textbf{Post information}}&\textbf{}\\ \hline \hline
\texttt{\textbf{Likes}} & Number of users who like this post.\\ 
\hline
\texttt{\textbf{Comments}}& Number of comments under this post.\\ 
 \hline
\texttt{\textbf{Create time}}&  \hspace{0.01cm} Creattion time of this post, format is in Unix \\ 
\texttt{} & time. \\ 
\hline
\texttt{\textbf{Caption}}& User description of this post. \\ 
\hline
\texttt{\textbf{Hashtag}}& Hashtag of this post. \\ 
\hline
\end{tabular}
\end{table}

Figure 1 shows one Instagram post consisting of pictures. 

\begin{figure}[ht!]
        \centering
        \subfigure[]{
                \includegraphics[ width=0.8\linewidth]{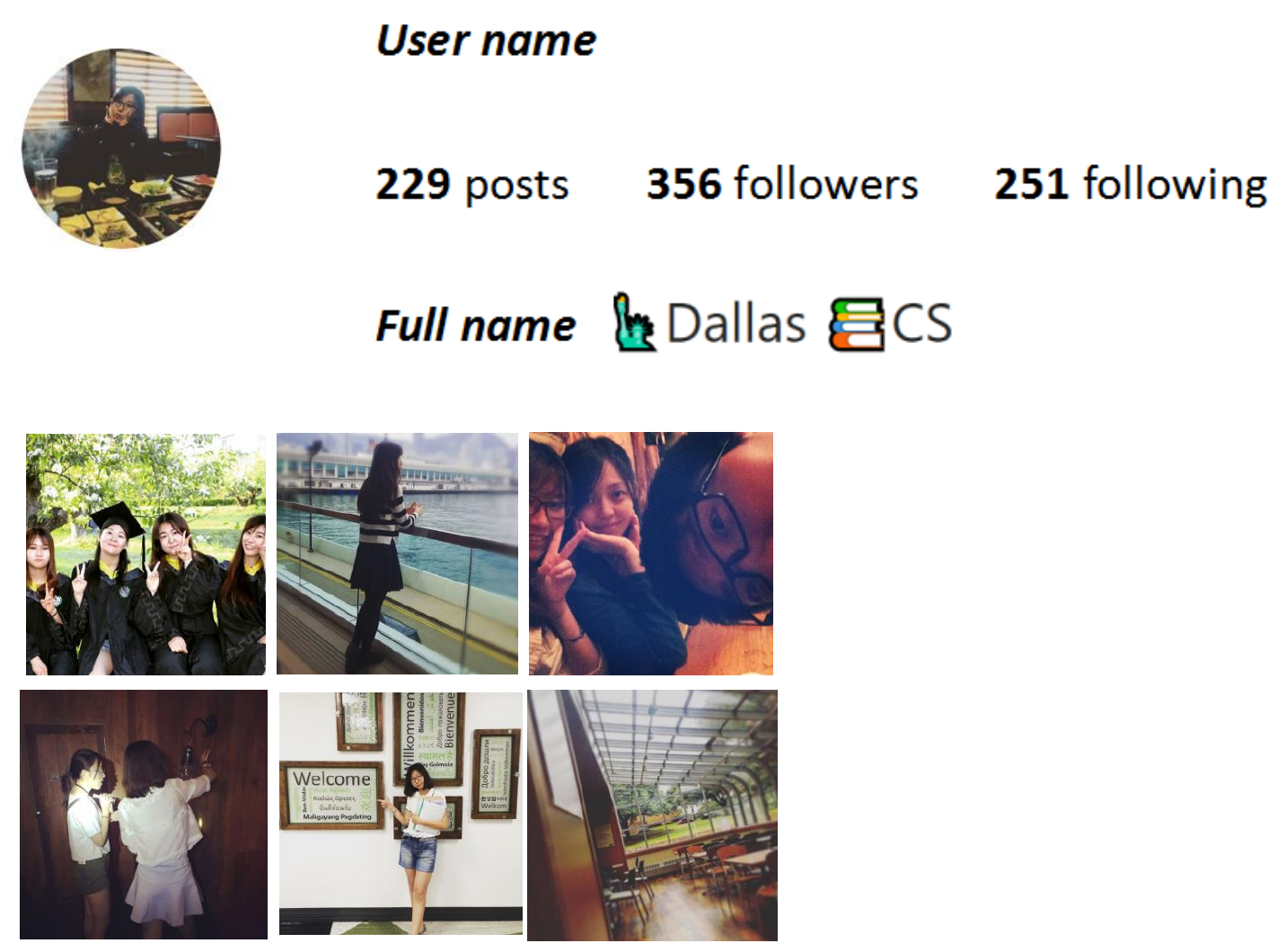}
               \label{fig:instagram_1}
        }
        \subfigure[]{
                \includegraphics[width=0.8\linewidth]{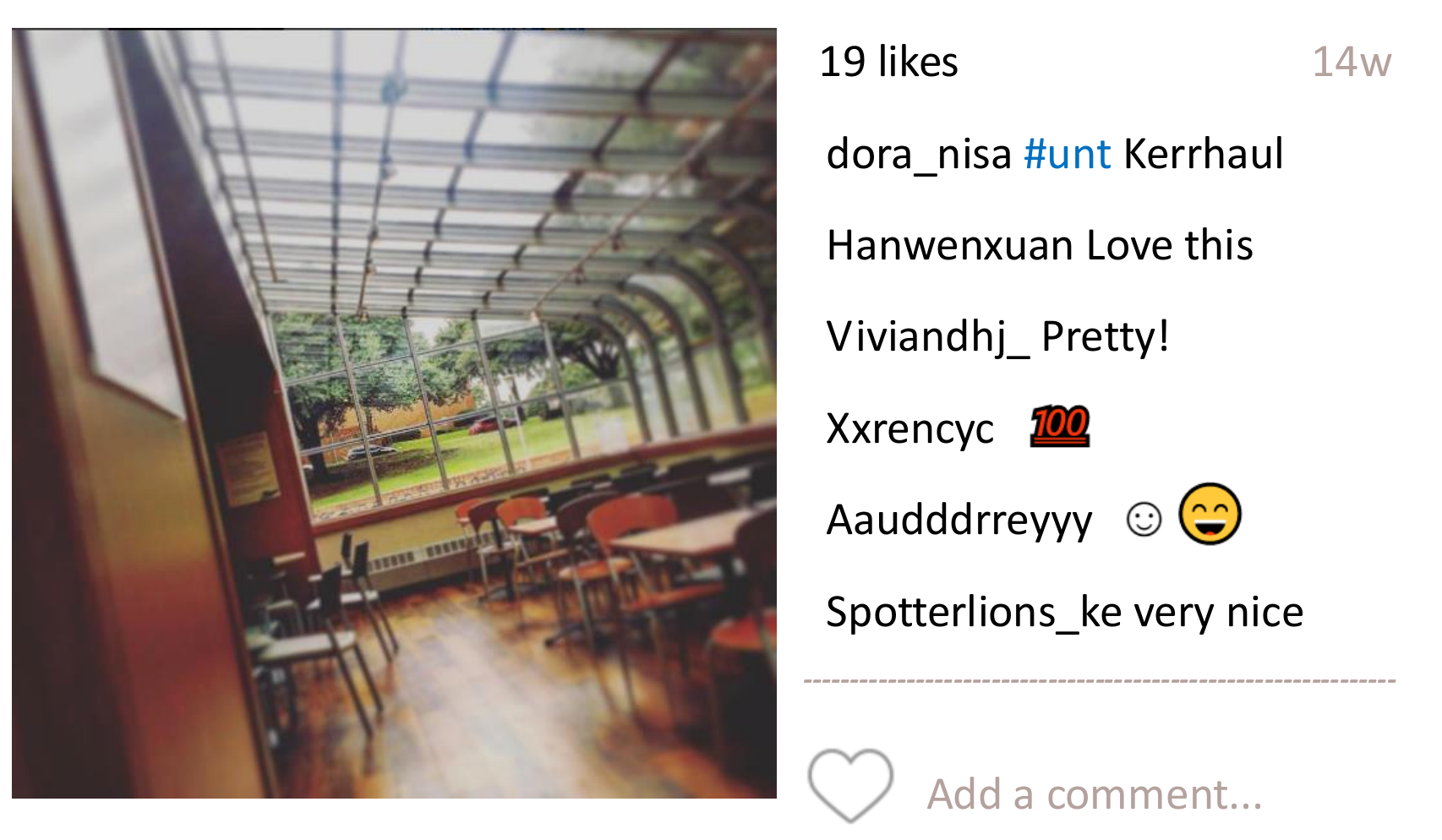}
               \label{fig:instagram_2}
        }
       
        \caption{User profile and a post in Instagram}
      
\end{figure}

\subsection{Factor Selection}
We first select \textit{followers}, \textit{following}, \textit{posts}, \textit{likes}, and \textit{comments} as factors. Posted pictures can also provide trust information. For each profile, we selected the 10 most recently posted pictures. For \textit{likes} and \textit{comments}, we sum up the number of \textit{likes} and number of \textit{comments} under the 10 posts of each user. We are interested in the influence of having people in the pictures on trust. We selected three picture attributes as factors. For the first factor, we counted the total number of people among these 10 pictures. For second factor, we counted the number of pictures that contained a person. For the third factor, we counted the number of pictures that contained the user themself. Together, we have 8 different factors from our dataset. 

\subsection{Trust Labeling}
There is limited availability of labeled data dealing with the relationship between social networks and rideshare systems - the two basic embodiments of our research. Survey is one of the most effective methods for generating ground truth dataset. We hosted a survey online by using Amazon Mechanical Turk. The content of each survey is based on the information from the selected Instagram public users that we collected. In the survey, it shows the profile picture of the user, total number of followers, total number of followings, and total number of posts. A user may have many posts. In our survey, we chose the 10 latest posts for each user. Each post contains detailed information, such as the caption, number of likes, number of comments, and the location of the post.

We asked respondents of the survey six questions: 

\begin{enumerate}[1)]
\item If this person were a driver for a rideshare service like Uber, I would trust this person to give me a ride.

Yes : I would trust. No: I would not trust.

\item If this person were the driver for a rideshare service like Uber, I would trust putting a close friend in a car with them.

Yes: I would trust. No: I would not trust.

\item Riding with this person would be more risky or less risky than average.

Yes: More risky. No: Less risky.

\item During the ride, I$'$d probably talk to this person more than average or less than average.

Yes: More than average. No: Less than average.

\item If this person was assigned to me as my rideshare partner, I$'$d feel safe riding in the car with them.

Yes: I$'$d feel safe. No: I$'$d not feel safe.

\item If you find out this person will be your rideshare partner, how likely are you to use a ride service.

Yes: More likely. No: Less likely.
\end{enumerate}

Mechanical Turk allows filtering of turkers according to certain criteria. In our survey, we randomly chose 5 different turkers to view each profile. Our criteria required each turker to have an approval rate above 98 percent and for him to be located in the United States. We analyzed the answers from 5 turkers for each profile entry. Table 4 shows the number of survey entries for each combination of numbers of \textquotedblleft Yes\textquotedblright and \textquotedblleft No\textquotedblright among the 5 turkers for each question. The ground truth trust labeling for each survey entry is chosen by majority vote.

\begin{table*}
\centering
\caption{The distribution of different percentages by answer \textquotedblleft Yes\textquotedblright or \textquotedblleft No\textquotedblright to the six questions}
\begin{tabular}{|c|c|c|c|c|c|c|} \hline
\textbf{}&\textbf{\textbf{Question 1}} &\textbf{\textbf{Question 2}} &\textbf{\textbf{Question 3}} &\textbf{\textbf{Question 4}} &\textbf{\textbf{Question 5}} &\textbf{\textbf{Question 6}}\\ \hline \hline
\texttt{\textbf{\textbf{Yes=0 N0=5}}} & 2 & 2 & 34 & 13 & 3 & 7 \\ 
\hline
\texttt{\textbf{\textbf{Yes=1 N0=4}}} & 8 & 11 & 19 & 17 & 7 & 13 \\ 
\hline
\texttt{\textbf{\textbf{Yes=2 N0=3}}} & 17 & 19 & 16 & 18 & 17 & 16 \\ 
\hline
\texttt{\textbf{\textbf{Yes=3 N0=2}}} & 14 & 13 & 15 & 17 & 13 & 14 \\ 
\hline
\texttt{\textbf{\textbf{Yes=4 N0=1}}} & 15 & 13 & 12 & 14 & 18 & 18 \\ 
\hline
\texttt{\textbf{\textbf{Yes=5 N0=0}}} & 44 & 42 & 4 & 21 & 42 & 32 \\ 
\hline
\end{tabular}
\end{table*}

\thispagestyle{plain}

\section{Factor Analysis}
Factor analysis is a statistical technique which extracts important factors from a variable group by reducing redundancy from a set of correlated variables. It explores the basic structure of observed data and presents the data structure with a few latent variables. These latent variables can reflect the most influential information of the original variables \cite{williams2010exploratory}. There are two main factor analysis methods. Exploratory Factor Analysis (EFA) and Confirmatory Factor Analysis (CFA) \cite{yong2013beginner}. The difference of these two methods is that EFA is used to uncover complex patterns by exploring the dataset and testing predictions; CFA is used to confirm hypotheses. In our case, we do not know the patterns, so we choose EFA as our factor analysis technique. In our analysis, there are eight trust factors. However, some of the factors may have some correlations. Therefore, we use EFA to reduce the factors to key factors. We performed five steps for the Exploratory Factor Analysis \cite{IBM}. First, we check whether the EFA approach suits our problem and dataset. Second, the extraction procedure is selected for the dataset. Third, method is decided to determine the total number of factors to retain. Fourth, we select the rotation method to yield a final solution. Fifth, a proper interpretation of each final factor is labeled. Failure to make a decision about one or more of the mentioned steps will cause erroneous results and limit the utility of EFA \cite{taherdoost2014exploratory}. Figure 2 shows the procedure of exploratory factor analysis. We chose the trust label of question 1 in the survey for factor analysis as it is the most general labeling of trust. 

\begin{figure*}[ht]
        \centering
        \includegraphics[width=0.8\linewidth]{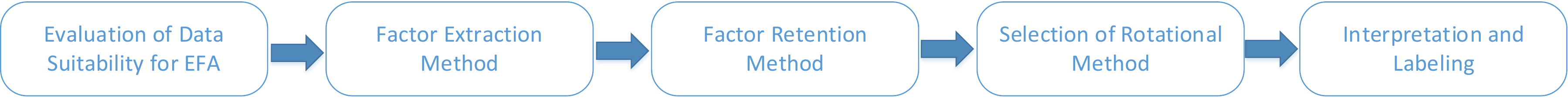}
        \caption{5-step Exploratory Factor Analysis}
        \label{fig:steps}
\end{figure*}

\subsection{Evaluation of data suitability}
There are several tests that can be conducted. The major tests include Kaiser-Meyer-Olkin (KMO) Measure of Sampling Adequacy and Bartlett$'$s Test of Sphericity. KMO measure varies between 0 and 1, and values closer to 1 are better. Table 5 shows that the KMO test value of our data is 0.714. Based on Tabachnick \cite{henson2006use}, 0.50 is considered suitable for factor analysis and Netemeyer \cite{netemeyer2003scaling} pointed that a KMO correlation above 0.60-0.70 is considered adequate for analyzing EFA output. So the KMO result indicates that our data is suitable for EFA. We use Bartlett$'$s Test of Sphericity to test the hypothesis that the correlation matrix is an identity matrix. The output should be significant (less than 0.05) for factor analysis to be suitable. The significant level of \textit{p} in our data is less than 0.001, which is much smaller than the indicator. It means the variables are correlated and have pattern relationships.

\begin{table}
\centering
\caption{KMO and Bartlett$'$s Test}
\begin{tabular}{|c|c|} \hline
\texttt{Kaiser-Meyer-Olkin Measure of Sampling Adequacy.} & .714 \\ 
\texttt{Bartlett$'$s Test of Sphericity Approx. Chi-Square}& 664.229\\
\texttt{                 df                             }& 28\\ 
\texttt{                 Sig.                           }& .0001\\
\hline
\end{tabular}
\end{table}

\subsection{Factor Extraction Methods}
A variety of methods have been developed to extract factors: Principle components method (PCA), Maximum likelihood method, principal axis factoring (PAF), unweighted least-squares method, generalized least squares method, alpha method and image factoring \cite{woolley2010evidence}. We chose PCA, the most commonly used method, as our factor extraction method. PCA is used to form linear combinations of the observed variables. The first component has maximum variance. Successive components explain progressively smaller portions of the variance and are uncorrelated with each other. PCA is used to obtain the initial factor solution \cite{thompson1996factor}. Table 6 provides the communality result. Communality is the proportion of each variable$'$s variance explained by the extracted factor structure. The results show that the community of each common extracted factor is nearly above 80\%, indicating the extracted common factor$'$s explanatory ability on each variable is strong. It indicates most information in the variables can be extracted by factors.

\begin{table}
\centering
\caption{Common factor variance}
\begin{tabular}{|c|c|c|} \hline
\textbf{}&\textbf{\textbf{Initial}}&\textbf{\textbf{Extraction}}\\ \hline \hline
\texttt{post} & 1.000 & .873\\ 
\texttt{follower}& 1.000 & .884\\
\texttt{following}& 1.000 & .895\\ 
\texttt{likes}& 1.000 & .861\\
\texttt{comments}& 1.000 & .786\\
\texttt{total\_person}& 1.000 & .763\\
\texttt{pic\_person}& 1.000 & .945\\
\texttt{self}& 1.000 & .918\\
\hline
\end{tabular}
\end{table}

\subsection{Factor Retention Methods}
Factor retention is an important step of exploratory factor analysis. Both under-extraction and over-extraction will have negative effects on the results. There are several criteria to decide how many factors to retain \cite{kaiser1960application}. Factor retention methods include: Kaiser$'$s criteria \cite{cattell1966scree}, Scree test \cite{hair2010multivariate}, and the cumulative percent of variance extracted. The results of these criteria do not always agree. Many researchers use multiple criteria to decide the number of factors to retain. In our factor retention steps, we checked all of these three criteria to make sure the final number of factors is convincible. First, we use Kaiser$'$s criteria, which is a rule of thumb. This criterion suggests to retain factors above the eigenvalue of 1. From Table 7, we can see that the factors are in the descending order based on the most explained variance. The initial eigenvalues show that in our case, there are three key factors with an eigenvalue greater than 1. So based on Kaiser$'$s criteria, we keep three factors. Second, we check the result from scree test. The scree test consists of factors and eigenvalues. For the scree test, we need to find the point at which the curve changes drastically. Figure 3 shows the result of the scree test. From point one to point three the curve is a steep slope suggesting to keep the first three points. It indicates the maximum number of components to retain is three. Finally, we used the cumulative percent of variance method to determine the number of factors to retain. For this method no fixed threshold exists, although certain percentages have been suggested. We select the threshold commonly used in humanities, which has the explained variance as low as 50-60\% \cite{kim1978factor}. In our case, we used rotated results to decide the number of key factors. From Table 7, the rotation sums of squared loadings indicates that the cumulative variance of the first three factors is 86.563\%, which is greater than 60\%. It means the first three factors are sufficient to represent the eight variables. 

\begin{table*}[ht!]
\caption{Total Variance}
\centering

\begin{tabular}{|c|r|c|c|c|c|c|c|c|c|}
    \hline
    \multirow{2}{*}{Component} 
    & \multicolumn{3}{|c|}{\textbf{Initial Eigenvalues}} & \multicolumn{3}{|c|}{\textbf{Extraction Sums of Squared Loadings}} & \multicolumn{3}{|c|}{\textbf{Rotation Sums of Squared Loadings}} \\ 
    \cline{2-10}
     & Total & \% of Variance & Cumulative \% & Total & \% of Variance & Cumulative \% & Total & \% of Variance & Cumulative \%\\
    \hline
    \hline
    \texttt{1} & 3.202 & 40.030 & 40.030 & 3.202 & 40.030 & 40.030 & 2.642 & 33.030 & 33.030\\
    \texttt{2} & 2.672 & 33.395 & 73.425 & 2.672 & 33.395 & 73.425 & 2.554 & 31.924 & 64.954\\
    \texttt{3} & 1.051 & 13.138 & 86.563 & 1.051 & 13.138 & 86.563 & 1.729 & 21.608 & 86.563\\
    \texttt{4} & .348 & 4.347 & 90.909 &  &  &  & &  & \\
    \texttt{5} & .324 & 4.048 & 94.957 &  &  &  & &  & \\
    \texttt{6} & .222 & 2.775 & 97.733 &  &  &  & &  & \\
    \texttt{7} & .145 & 1.818 & 99.550 &  &  &  & &  & \\
    \texttt{8} & .036 & .450 & 100.000 &  &  &  & &  & \\
    \hline
    \end{tabular}%
\label{tab:variance}
\end{table*}

\begin{figure}[ht]
        \centering
        \includegraphics[width=0.8\linewidth]{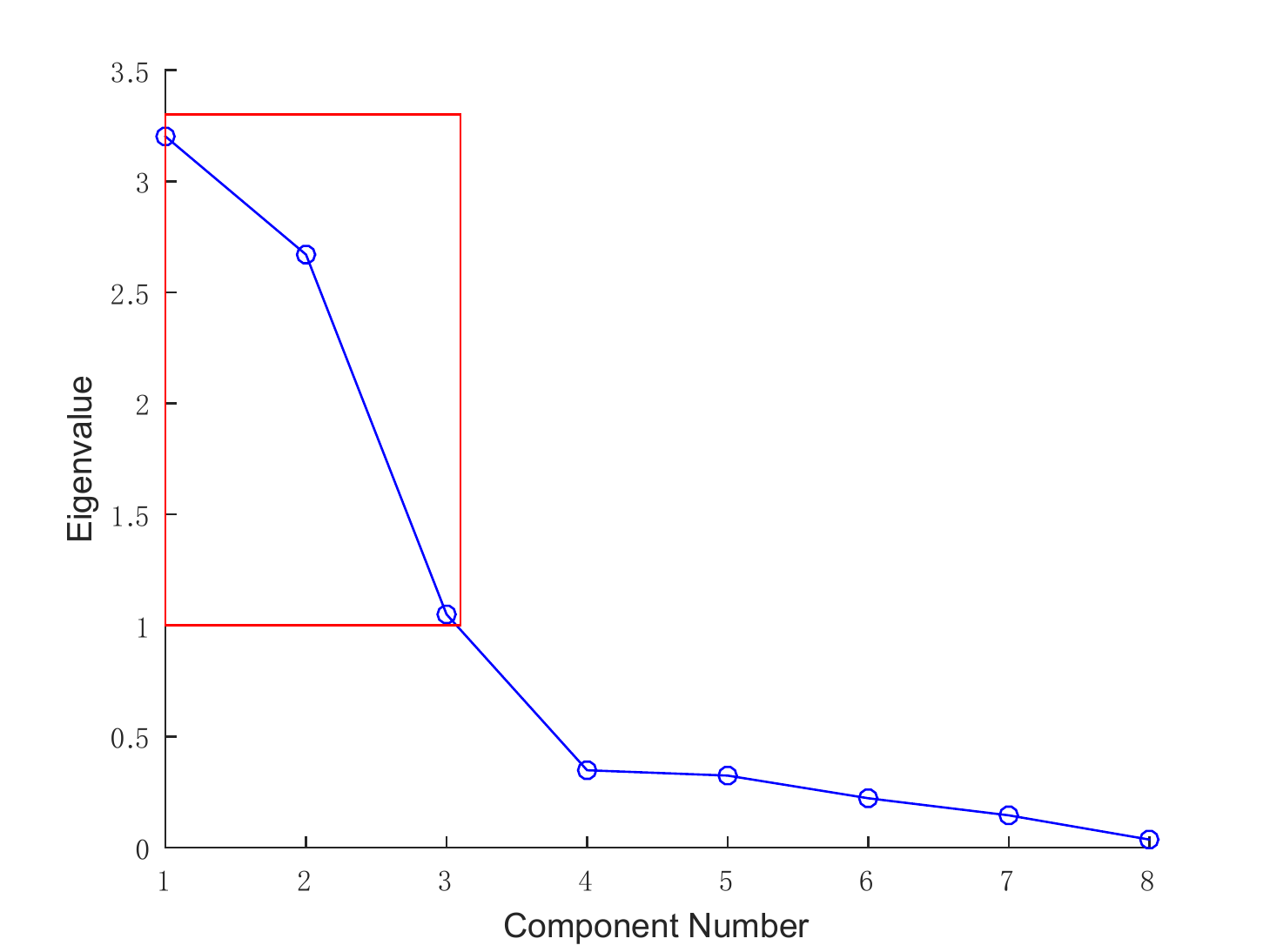}
        \caption{Scree Plot}
        \label{fig:overview}
\end{figure}

\subsection{Rotation Methods}
Rotation methods deal with a variable that is related to more than one factor. The key idea of rotation is that it maximizes high-item loadings and minimizes low-item loadings. By doing this, it will be much easier to facilitate interpretation of the factors. Table 8 shows the factor matrix, which is the factor loadings before rotation, and Table 9 shows the rotated factor loadings. In our study we use orthogonal varimax approach for our rotation method. Comparing Table 8 and Table 9 we can find that, after the rotation step, the approach can produce factor structures that are uncorrelated. From Table 9 we can see that rotation and suppressing small coefficients helps with the interpretation. In the rotated matrix, we can easily assign each variable to the corresponding factor without overlapping. In our case, we choose a loading cutoff of 0.36.

\thispagestyle{plain}

\subsection{Interpretation and Labeling}
In this step, we need to examine which variables are attributing to a factor and then give the factor a name. For instance, if you have a factor the name \textquoteleft emotional\textquoteright, and variables that load high on this factor are \textquoteleft laugh\textquoteright, \textquoteleft angry\textquoteright, and \textquoteleft crying\textquoteright, you also need to make sure that an unrelated variable such as \textquoteleft writing\textquoteright does not load onto this factor. To give a meaningful interpretation, usually two or three variables must load on a factor. The labeling of a factor is a theoretical and subjective process. In our case, after varimax rotation, we had three significant factors and assigned each variable to its corresponding factor. From Table 9, we can see that, factor 1 consists of total\_person, pic\_person, and self. Factor 2 contains follower, likes, and comments. The third factor contains following and post. 

\begin{description}
\item[Factor 1:] Social Proof

This factor, Social Proof, represents the positive influence on trust of users providing evidence, or proof, that they are social. Pictures of the user and other people offer \textquoteleft proof\textquoteright that the user is sociable and likable. This factor contains the total number of people visible in the 10 pictures, the total number of pictures (out of 10) that contained a person, and the total number of pictures that contained the user personally.

\item[Factor 2:] Social Approval

Social Approval by other people on activity of the user$'$s account, expressed through \textquoteleft like\textquoteright of pictures and comments on the user$'$s profile, increased perceptions of trust and reduced perceptions of risk among potential rideshare partners. This factor illustrates the positive influence of social approval by other people on respondents$'$ perceptions of trust.

\item[Factor 3:] User Self-Disclosure

Self-disclosures by the user increased perceptions of user trust. A higher number of total posts by the user was associated with increased trust, as was the total number of people that the user \textquoteleft follows\textquoteright on their accounts. These user self-disclosures of total posts shared and total number of accounts the user follows, illustrate openness and sociability of the user, factors that increase trust and reduce risk.

\end{description}

\begin{table}[ht!]
\caption{Factor Matrix}
\centering

\begin{tabular}{|c|r|c|c|}
    \hline
    \multirow{2}{*}{} 
    & \multicolumn{3}{|c|}{\textbf{Component}} \\ 
    \cline{2-4}
     & 1 & 2 & 3 \\
    \hline
    \hline
    \texttt{post} & .760 & -.106 & -.532\\ 
    \texttt{follower}& .847 & -.288 & .288\\
    \texttt{following}& .676 & .115 & -.652\\ 
    \texttt{likes}& .763 & -.440 & .293\\
    \texttt{comments}& .747 & -.334 & .341\\
    \texttt{total\_person}& .251 & .817 & .180\\
    \texttt{pic\_person}& .365 & .894 & .111\\
    \texttt{self}& .338 & .889 & .110\\
    \hline
    \end{tabular}%
\label{tab:factor_matrix}
\end{table}

\begin{table}[ht!]
\caption{Factor Matrix after Varimax rotation}
\centering

\begin{tabular}{|c|r|c|c|}
    \hline
    \multirow{2}{*}{} 
    & \multicolumn{3}{|c|}{\textbf{Component}} \\ 
    \cline{2-4}
     & 1 & 2 & 3 \\
    \hline
    \hline
    \texttt{post} & .018 & .350 & \cellcolor[rgb]{.7,.9,.9}.886\\ 
    \texttt{follower}& .059 & \cellcolor[rgb]{.7,.9,.9}.909 & .234\\
    \texttt{following}& .170 & .139 & \cellcolor[rgb]{.7,.9,.9}.920\\ 
    \texttt{likes}& -.107 & \cellcolor[rgb]{.7,.9,.9}.904 & .181\\
    \texttt{comments}& -.003 & \cellcolor[rgb]{.7,.9,.9}.876 & .133\\
    \texttt{total\_person}& \cellcolor[rgb]{.7,.9,.9}.873 & -.025 & .001\\
    \texttt{pic\_person}& \cellcolor[rgb]{.7,.9,.9}.965 & -.002 & .123\\
    \texttt{self}& \cellcolor[rgb]{.7,.9,.9}.952 & -.021 & .109\\
    \hline
    \end{tabular}%
\label{tab:factor_rotation_matrix}
\end{table}

Among the three factors, Social Proof is associated with persons in pictures and influenced trust formation the most. Pictures may be manipulated to increase trustworthiness. However, technical advances in picture analysis using deep learning makes recognizing a person very accurate. It is not difficult to require a real profile picture and identify and label each person in the picture. Social Approval, on the other hand, is controlled by other users and difficult to manipulate. User self-disclosure explains 13\% of the variance and will take sustained efforts over time to arbitrarily boost it up.

\section{Results of Logistic Regression}
In our survey, we have six different questions. The answer to each of our survey questions is binary. Thus, we built a logistic regression model to analyze the relation between factors and the results of surveys. We used precision, recall, and F-measure as the indicators of logistic regression. We first analyzed the original eight factors. Table 10 shows detailed results. Then, we analyzed the three factors produced by factor analysis. The detailed results are shown in table 11.

\begin{table*}
\centering
\caption{Result of eight factors}
\begin{tabular}{|c|c|c|c|c|c|c|} \hline
\textbf{}&\textbf{\textbf{Question 1}} &\textbf{\textbf{Question 2}} &\textbf{\textbf{Question 3}} &\textbf{\textbf{Question 4}} &\textbf{\textbf{Question 5}} &\textbf{\textbf{Question 6}}\\ \hline \hline
\texttt{\textbf{\textbf{Precision}}} & 0.783 & 0.782 & 0.714 & 0.609 & 0.746 & 0.700 \\ 
\hline
\texttt{\textbf{\textbf{Recall}}} & 0.790 & 0.788 & 0.730 & 0.610 & 0.760 & 0.710 \\ 
\hline
\texttt{\textbf{\textbf{F-Measure}}} & 0.786 & 0.782 & 0.715 & 0.609 & 0.750 & 0.699 \\ 
\hline
\end{tabular}
\end{table*}

\begin{table*}
\centering
\caption{Result of three factors}
\begin{tabular}{|c|c|c|c|c|c|c|} \hline
\textbf{}&\textbf{\textbf{Question 1}} &\textbf{\textbf{Question 2}} &\textbf{\textbf{Question 3}} &\textbf{\textbf{Question 4}} &\textbf{\textbf{Question 5}} &\textbf{\textbf{Question 6}}\\ \hline \hline
\texttt{\textbf{\textbf{Precision}}} & 0.846 & 0.843 & 0.797 & 0.661 & 0.823 & 0.726 \\ 
\hline
\texttt{\textbf{\textbf{Recall}}} & 0.904 & 0.868 & 0.913 & 0.750 & 0.890 & 0.828 \\ 
\hline
\texttt{\textbf{\textbf{F-Measure}}} & 0.874 & 0.855 & 0.851 & 0.703 & 0.855 & 0.774 \\ 
\hline
\end{tabular}
\end{table*}

From Table 10 and Table 11, we can see that, the three key factors have a better result than the original eight factors.

\begin{figure}[!ht]
        \centering
        \includegraphics[width=0.8\linewidth]{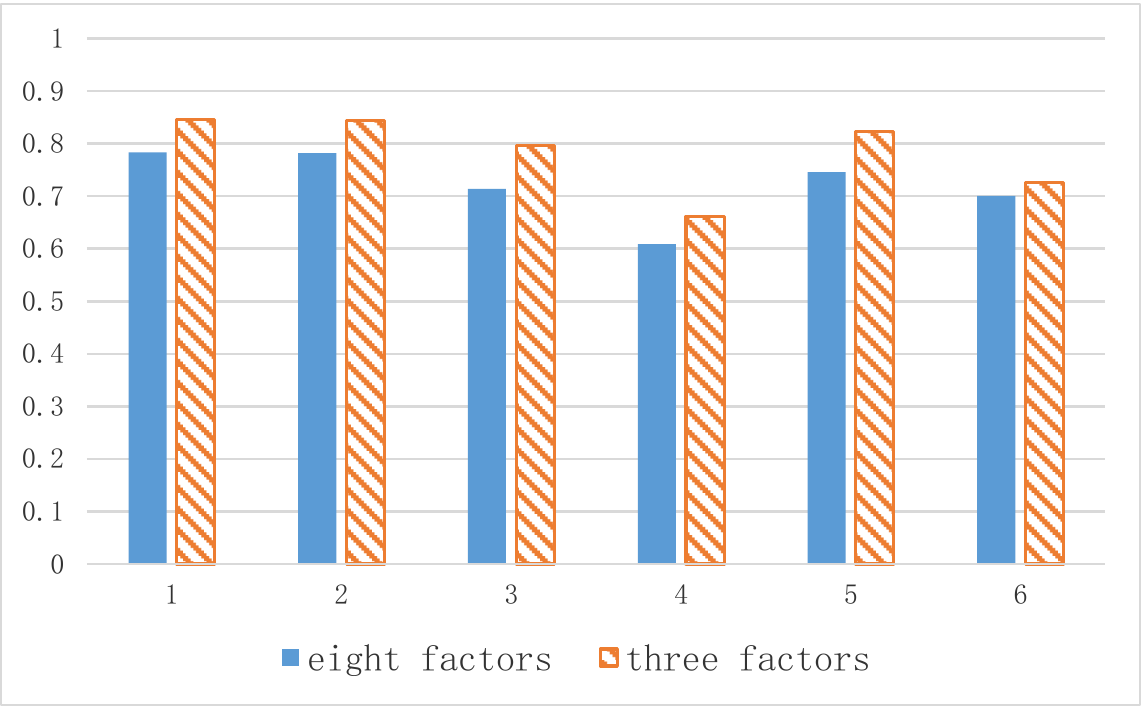}
        \caption{Precision}
        \label{fig:precision}
\end{figure}

\begin{figure}[!ht]
        \centering
        \includegraphics[width=0.8\linewidth]{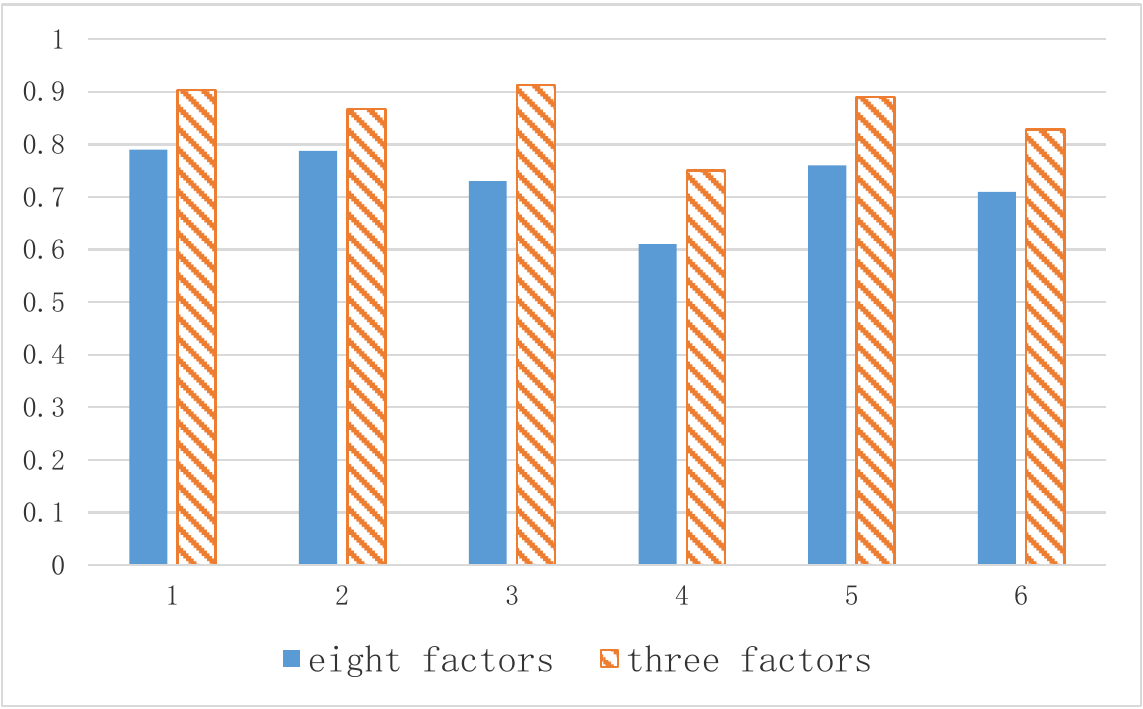}
        \caption{Recall}
        \label{fig:recall}
\end{figure}

\begin{figure}[!ht]
        \centering
        \includegraphics[width=0.8\linewidth]{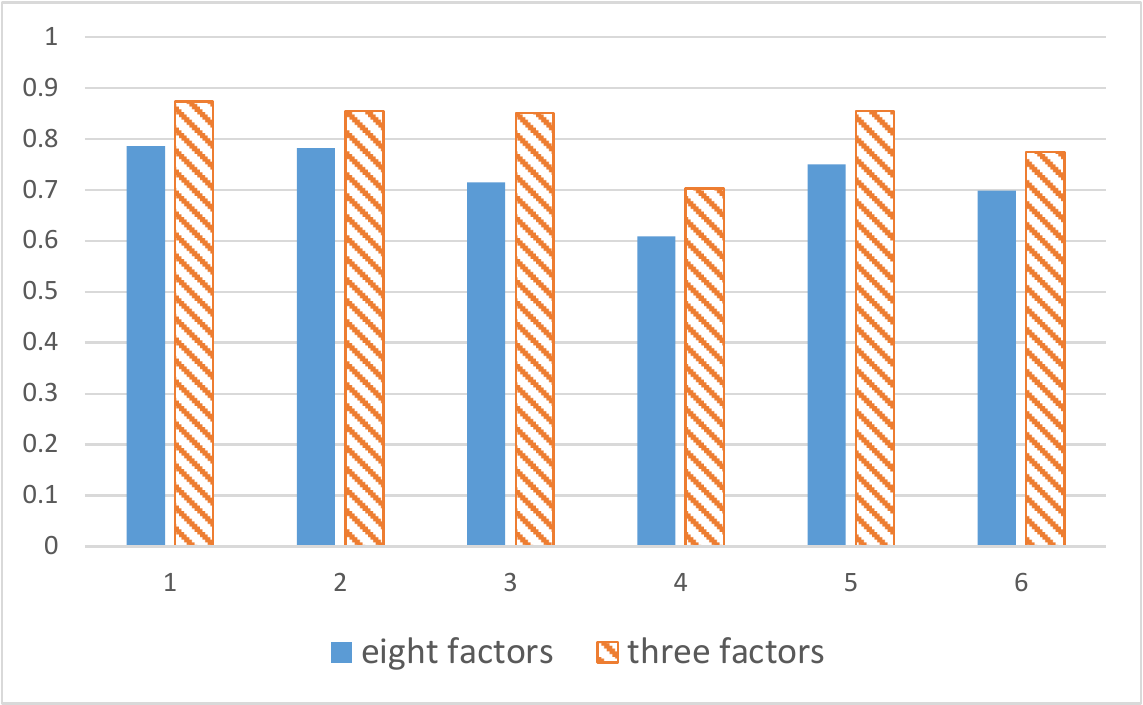}
        \caption{F-Measure}
        \label{fig:fmeasure}
\end{figure}

Figure 4 to Figure 6 also shows the comparison of eight factors and three key factors. We can see obviously that the result of three key factors is better than the result of eight factors in precision, recall, and F-measure, among all of the six questions.

\section{Conclusion}
In this paper, we combined social media and ridesharing systems together to analyze the influence of trust on the potential usage of ridesharing services. In our work, we collected the detailed social information from Instagram. We used Mechanical Turk to conduct a survey to learn the relationship between trust factors and the use of ridesharing. By analyzing different social attributes, we chose eight social attributes as trust factors. We then used EFA to summarize the eight factors into three key factors. Finally, we built a logistic regression model by using the results from the survey with eight factors and three factors separately. The results show that with using eight factors, we can achieve 78.6\% in F-Measure. By using only the three most important factors, the result will be improved to 87.4\%.

\bibliographystyle{ACM-Reference-Format}
\bibliography{sigproc} 

\thispagestyle{plain}

\end{document}